\documentclass[iop]{emulateapj}

\newcommand{\pwn}{HESS J1640$-$465}
\newcommand{\snr}{G338.3$-$0.0}
\newcommand{\chandra}{{\em Chandra}}

\newcommand{\fermi}{{\em Fermi}}
\newcommand{\xmm}{{\em XMM}}

\usepackage{natbib,graphicx}

\slugcomment{Accepted for Publication in ApJ}

\shorttitle{Fermi Observations of HESS J1640$-$465}
\shortauthors{Slane et al.}

\begin{document}

\title{Fermi Detection of the Pulsar Wind Nebula HESS J1640$-$465}

\author{P.~Slane\altaffilmark{1}, D.~Castro\altaffilmark{1,2},
S.~Funk\altaffilmark{3}, Y.~Uchiyama\altaffilmark{3},
A.~Lemiere\altaffilmark{4}, J.~D.~Gelfand\altaffilmark{5,6}, \&
M.~Lemoine-Goumard\altaffilmark{7}}

\altaffiltext{1}{Harvard-Smithsonian Center for Astrophysics, 60 Garden
Street, Cambridge, MA 02138, USA; slane@cfa.harvard.edu;
dcastro@head.cfa.harvard.edu}
\altaffiltext{2}{
Departamento de F\'isica, Universidad Sim\'{o}n Bol{\'\i}var, Valle de
Sartenejas, Venezuela}
\altaffiltext{3}{Kavli Institute for Particle Astrophysics and Cosmology,
Stanford Linear Accelerator Center, Stanford, CA 94025, USA;
funk@slac.stanford.edu; uchiyama@slac.stanford.edu}
\altaffiltext{4}{Institut de Physique Nucleaire, IN2P3/CNRS, 15 rue Georges
Clemenceau, 91400 Orsay, France}
\altaffiltext{5}{Center for Cosmology and Particle Physics, New York University,
Meyer Hall of Physics, 4 Washington Place, New York, NY 10003, USA}
\altaffiltext{6}{NSF Astronomy and Astrophysics Postdoctoral Fellow}
\altaffiltext{7}{Universit\'e de Bordeaux, Centre d'\'Etudes Nucl\'eaires 
Bordeaux Gradignan, CNRS-IN2P3, UMR 5797, Gradignan 33175, France}

\begin{abstract}
We present observations of \pwn\ with the \fermi-LAT. The source is detected
with high confidence as an emitter of high-energy gamma-rays. The spectrum
lacks any evidence for the characteristic cutoff associated with emission
from pulsars, indicating that the emission arises primarily from the
pulsar wind nebula. Broadband modeling implies an evolved nebula with
a low magnetic field resulting in a high $\gamma$-ray to X-ray flux ratio.
The \fermi\ emission exceeds predictions of the broadband model, and has
a steeper spectrum, possibly resulting from a distinct excess of low
energy electrons similar to what is inferred for both the Vela X and
Crab pulsar wind nebulae.
\end{abstract}


\section{Introduction}

The basic structure of a pulsar wind nebula (PWN) is determined by the
spin-down energy injected by the central pulsar and the interaction
of the nebula with the interior regions of the supernova remnant
(SNR) in which it evolves. Losses from synchrotron radiation in the
nebular magnetic field, whose strength depends both on the nature
of the injected wind and on the evolving size of the PWN, inverse-Compton
scattering of ambient photons by the energetic electron population
within the nebula, and adiabatic expansion as the nebula sweeps up
the surrounding supernova ejecta, all combine to determine emission
structure and long-term evolution of the nebula. (See Gaensler \&
Slane 2006 for a review.) Multiwavelength observations of PWNe
provide crucial information on the underlying particle spectrum
which, in turn, strongly constrains both the magnetic field strength
and the stage of evolution. Of particular importance is the spectrum
of particles injected into the PWN. While this is typically assumed
to be a power law, it has long been known that the spectrum of the
Crab Nebula is not fully consistent with such a structure; changes
in the spectral index in the optical band seem to indicate inherent
features in the injection spectrum, and the large population of
radio-emitting particles may have a completely distinct origin from
that of the higher energy particles.

Complex structure in the observed spectrum of a PWN can originate
in a number of ways that are associated with the long-term evolution,
including synchrotron losses and interaction with the SNR reverse
shock (Reynolds \& Chevalier 1984; Gelfand, Slane, \& Zhang 2009).
In addition, recent studies of the spectrum immediately downstream of
the wind termination shock show that, at least in some cases, the
injection spectrum itself deviates significantly from a simple power
law (Slane et al.  2008), and particle-in-cell simulations of the
acceleration process produce spectra that are well-described by a
Maxwellian population with a power law tail (Spitkovsky 2008).  Any
such structure in the spectrum of the injected particles imprints
itself on the broadband emission spectrum of the entire nebula. The
resulting breaks or regions of curvature in the emission spectrum,
and the frequencies at which they are observed, depend upon the
energy at which features appear in the electron spectrum as well
as the means by which the photons are produced (e.g. synchrotron
radiation or inverse-Compton emission).  To fully understand the
nature of the particle injection, as well as the long-term evolution
of PWNe, it is thus crucial to study the emission structure over
the entire electromagnetic spectrum.

Gamma-ray observations have provided an important window into the
late-phase structure of PWNe. While the X-ray emission from older
PWNe is relatively weak due to the long-term decline in magnetic
field strength, inverse-Compton scattering of the cosmic microwave
background (CMB), as well as other ambient photons, by the energetic
particles in the nebula produce very energetic photons, providing
unique discovery space for these systems. Even in younger nebulae,
the $\gamma$-ray emission provides a crucial probe of the particle
spectrum in energy regions that are not accessible with other
observations. The inverse-Compton emission from photons with a
characteristic temperature $T$ peaks at an energy $\epsilon_{ic}
\approx 5 \gamma^2 kT$ for scattering from electrons of energy
$\gamma m_e c^2$.  CMB photons scattered into the bandpass of the
\fermi\ Large Area Telescope (LAT), for example, originate from
interactions with electrons in the approximate energy range 0.25
-- 4 TeV. Such particles produce synchrotron radiation with photon
energies in the range $\epsilon_{s} \approx 0.03 - 7 B_{10}$~eV,
where $B_{10}$ is the magnetic field strength in units of 10$\mu$G.
Depending on the magnetic field, which can range from $>100 \mu$G
or higher for young PWNe, down to $\sim 5 \mu$G for highly evolved
systems, the synchrotron emission from particles in this energy
range can be difficult to detect due to instrumentation limitations
(at long radio wavelengths), absorption (in the optical/UV range)
or confusion with the bright sky Galactic background (in the
infrared).  \fermi\ measurements can thus provide a unique probe
of the emission from a significant population of the particle
spectrum in PWNe.

\pwn\ (see Figure 1) is an extended source of very high energy
$\gamma$-ray emission discovered with the High Energy Stereoscopic
System (H.E.S.S.) during a survey of the Galactic plane (Aharonian
et al.  2006).  Centered within the radio SNR G338.3$-$0.0 (Shaver
\& Goss 1970), the deconvolved TeV image of the source has an RMS
width of $2.7 \pm 0.5$~arcmin (Funk et al. 2007). HI measurements
show absorption against G338.3$-$0.0 out to velocities corresponding
to the tangent point, indicating a distance of at least 8 kpc
(Lemiere et al. 2009), and thus implying a rather large size for
the PWN ($R_{PWN} > 6.4 d_{10} $~pc, where $d_{10}$ is the distance
in units of 10~kpc).  X-ray observations with \xmm\ (Funk et al.
2007) and Chandra (Lemiere et al. 2009) establish the presence of
an accompanying X-ray nebula and an X-ray point source that appears
to be the associated neutron star. In addition, Aharonian et al.
(2006) noted the presence of the unidentified {\em EGRET} source
3EG~J1639$-$4702 located at a nominal position 34$^\prime$ from
\pwn, but the very large error circle on its position made any
association with \pwn\ highly uncertain.  Here we report on
observations of \pwn\ with the \fermi-LAT.  The observations and
data analysis are summarized in Section 2, and a discussion of the
observed $\gamma$-ray emission is discussed in the context of the
evolutionary state of \pwn\ in Section 3. Our conclusions are
summarized in Section 4.

\begin{figure}[t]
\epsscale{1.15}
\plotone{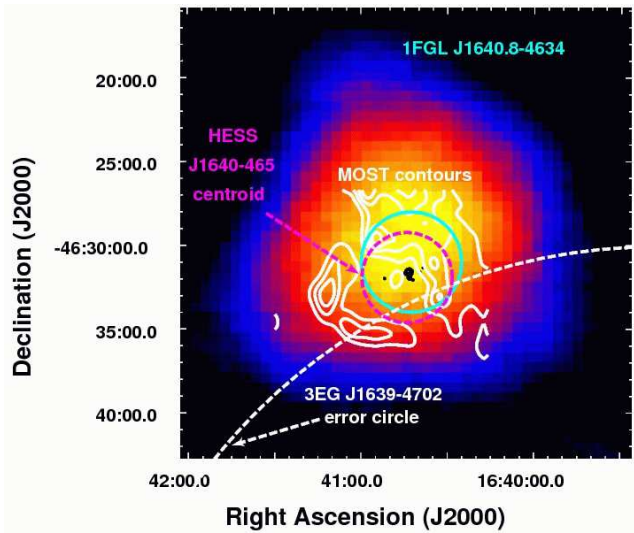}
\epsscale{1.0}
\caption{
\fermi\ LAT image (2 - 200 GeV) of \pwn. The cyan circle indicates
the uncertainty in the centroid of the \fermi\ LAT source, the
magenta circle indicates the 95\% encircled flux distribution of
the HESS image, and the white circle indicates the 95\% probability
contour for the position of 3EG J1639$-$4702. The white contours
outline radio emission from G338.3$-$0.0 while the black contours
at the center outline extended X-ray emission observed with \xmm.
A compact X-ray source detected with \chandra\ resides within the
X-ray contours.
}
\end{figure}

\section{Observations and Data Analysis}

We investigated $\gamma$-ray events acquired from the region
surrounding \pwn\ with the \fermi-LAT during the period 2008 August
5 to 2009 November 13. Standard event selection was applied, using
the ``diffuse'' class as defined in Atwood et al. (2009), using
events with zenith angles less than 105 degrees to minimize the
portion of the Earth limb in the LAT field-of-view (Abdo et al.
2009a). The ``Pass6 version 3'' instrument response functions were
used.  Standard analysis tools available from the \fermi\ Science
Support Center (version v9r15p2) were used for the reduction and
analysis.  In this work, the mapcube file gll\_iem\_v02.fit is used
to describe the Galactic $\gamma$-ray emission, and the isotropic
component is modeled using the isotropic\_iem\_v02.txt table.  Data
analysis details follow those in Castro \& Slane (2010).

For source detection and spatial analysis of the source, we used
only events with energies in the range 2 - 200 GeV converting in
the front section of the tracker, where the Point Spread Function
(PSF) is narrower. At 2 GeV, the 68\% containment radius of the PSF
is approximately 18 arcmin, which is considerably larger than the
source of interest.  Based on an unbinned maximum likelihood analysis,
using the {\tt gtlike} routine, a LAT source coincident with \pwn\
is detected with a significance of $\sim 17 \sigma$ based on the
TS statistic (Mattox et al. 1996).  The counts map is shown in
Figure 1, along with the uncertainty in the centroid position based
on our analysis (cyan circle). We have also included the HESS error
circle (magenta) and contours from the \xmm\ observation of \pwn\
(black) and the MOST observation of \snr (white contours). The error
circle for 3EG~J1639$-$4702 is indicated by a dashed white
circle.\footnote{EGRET position contours are not necessarily circular;
the radius of the contour shown, from Hartman et al. (1999), contains
the same solid angle as the formal 95\% contour.} The centroid of
the LAT emission is located at $16^{\rm h} 40^{\rm m} 46^{\rm s}$,
$-46^\circ 30^\prime 44^{\prime\prime}$, in good agreement with the
position of \pwn, and the brightness distribution is consistent
with an unresolved source (Figure 2); models for a disk of extent
between $0.05 - 0.2$~degrees significantly degrade the quality of
the fit, providing strong evidence for an unresolved source.

\begin{figure}[t]
\epsscale{1.20}
\plotone{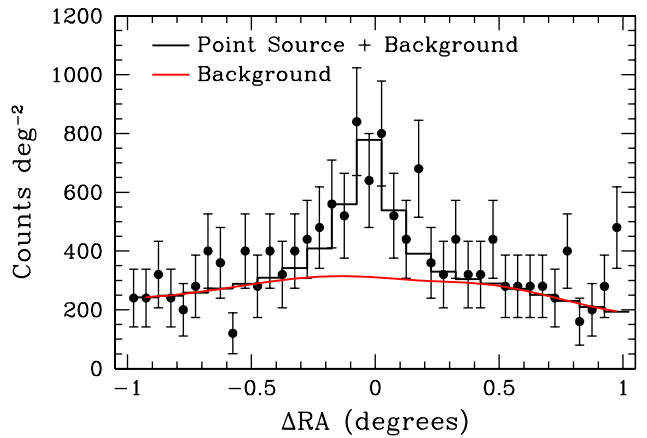}
\epsscale{1.00}
\caption{
Profile of the \fermi\ LAT emission from \pwn. The histogram
corresponds to the best-fit point source profile, including the
diffuse Galactic and extragalactic background which is shown separately
in red.
}
\end{figure}

The source spectrum was extracted using both front and back events,
and covering the energy range $0.2 - 51.2$~GeV.  Standard background
models were used to account for both Galactic and extragalactic
diffuse emission as well as instrumental background.  Contributions
from field sources identified in the one-year \fermi-LAT First
Source Catalog (Abdo et al. 2010a) were included in the analysis.
The LAT spectrum for \pwn\ is shown in Figure 3, and is well-described
by a power law with $\Gamma = 2.30 \pm 0.09$ and a $F(>100{\rm\
MeV}) = (2.8 \pm 0.4) \times 10^{-7}{\rm\ photons\ cm^{-2}}{\rm\
s}^{-1}$ [or $(1.7 \pm 0.2)) \times 10^{-10}{\rm\ erg\ cm^{-2}\
s}^{-1}$ in the $0.1 - 300$~GeV band] based on spectral fits for
which statistical error and systematic errors associated with
uncertainties in the Galactic background level (estimated by
artificially changing the Galactic background normalization by $\pm
3\%$ -- see Abdo et al. 2009b) are added in quadrature.

We note that \fermi\ observations of pulsars reveal spectra with
distinct cutoffs above $\sim 1 - 10$~GeV (e.g. Abdo et al. 2009).
The absence of such a cutoff in Figure 3 implies that that the bulk
of the emission does not arise directly from an unseen pulsar in
\pwn.  Our fits imply a lower limit of 40~GeV (at the 90\% confidence
level) for any exponential cutoff energy in the spectrum.  Joint
fits with the HESS spectrum are also well-described by a power law.
Addition of a second power law, with an exponential cutoff, does
not improve the fit. Such a fit can, however, accommodate $\sim
20\%$ of the observed flux in the second power law for cutoff
energies between $1 - 8$~GeV (beyond which such a component is
statistically disfavored).

\begin{figure}[t]
\epsscale{1.2}
\plotone{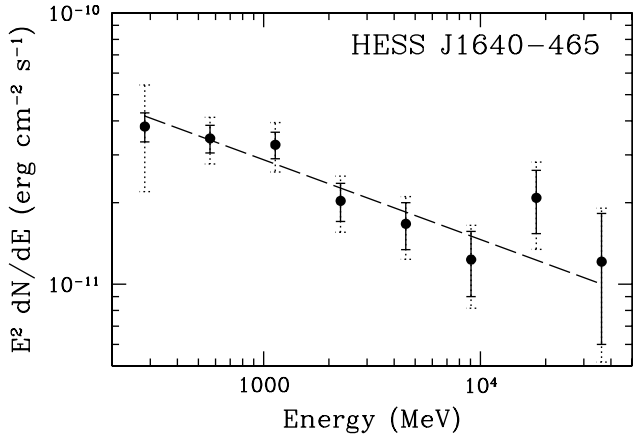}
\epsscale{1.0}
\caption{
\fermi\ LAT spectrum of \pwn. Statistical (systematic) uncertainties are
indicated by solid (dashed) error bars. The dashed line corresponds to
the best-fit power law model described in the text.
}
\end{figure}

The source 1FGL~J1640.8$-$4634, from the one-year catalog, is
coincident with the source we detect, and the quoted flux is in
agreement with our measurements as well.  The flux of the source
is consistent with that of 3EG~J1639$-$4702; the spectral index is
somewhat flatter, though also consistent within the uncertainties.

\section{Discussion}

The evolutionary state of a composite supernova remnant system is
strongly constrained by the observed size of the SNR and the inferred
spin-down properties of the associated pulsar. Radio observations
establish a radius $R_{SNR} \sim 11.6 d_{10}$~pc for \snr.  The
observed extent of \pwn\ constrains the radius of the PWN to $R_{PWN}
> 6.4 d_{10}$~pc.\footnote{Here, and throughout, $R_{SNR}$ refers to
the distance from the SNR center to the outer edge of its blast wave,
while $R_{PSR}$ refers to the distance from the pulsar to the outer
boundary of its wind nebula.}
For evolution in the Sedov phase, the SNR radius
is 
\begin{equation} R_{SNR} = 4.9 \left(\frac{E_{51}}{n_0}\right)^{1/5}
t_3^{2/5}{\rm\ pc} \end{equation} 
where $E_{51}$ is the supernova explosion energy in units of
$10^{51}$~erg, $n_0$ is the number density of the ambient medium,
and $t_3$ is the SNR age in kyr. The evolution of the PWN radius
through the SNR ejecta is given by (Chevalier 1977) 
\begin{equation}
R_{PWN} \sim 1.1 \dot{E}_{0,38}^{1/5} E_{51}^{3/10} M_{10}^{-1/2}
t_3^{6/5} {\rm\ pc} \end{equation} 
where $\dot{E}_{0,38}$ is the initial spin-down power in units of
$10^{38}{\rm\ erg\ s}^{-1}$ and $M_{10}$ is the ejecta mass in units
of $10 M_\odot$. Figure 4 illustrates the evolution of the SNR and
PWN radii under such assumptions for a range of values for the
ambient density and initial spin-down power, assuming an ejecta
mass of $8 M_\odot$ and $E_{51} = 1$. For the observed radius of
\snr, we see that the age must be around $5 - 8$~kyr for reasonable
ambient conditions.  The PWN has ideally expanded to a larger radius
by this time, which means that under real conditions the SNR reverse
shock has already encountered and disrupted the PWN.  This is
illustrated in Figure 4 where the solid blue and red curves show
the SNR and PWN radii (respectively) as a function of time for the
model shown in Gelfand et al. (2009) with $\dot{E_0} = 10^{40} {\rm\
erg\ s}^{-1}$, $M_{ej} = 8 M_\odot$, $n_0 = 0.1 {\rm\ cm}^{-3}$,
and $E_{51} = 1$.  Here the early SNR evolution is calculated using
the solution from Truelove \& McKee (1999), and the SNR and PWN
evolution is treated self-consistently.  The SNR curve initially
rises more quickly than the dashed blue curves, which assume a Sedov
solution from the onset, but approaches the Sedov solution at later
times. The solid red curve shows a distinct reduction in the radius
of the PWN upon compression by the SNR reverse shock. Such a reverse
shock interaction is consistent with both the inferred size of \pwn\
and with the observed offset between the putative pulsar and the
surrounding nebula (Funk et al.  2007, Lemiere et al. 2009).

A full exploration of the parameter space constrained by both the
spatial and spectral properties of \pwn\ and \snr\ is beyond the scope of
this initial investigation. Here we have modeled the PWN emission
following the description in Lemiere et al. (2009). A simple power
law injection of particles from the pulsar into a 1-zone nebula is
assumed, and radiative losses are assumed to be dominated by
synchrotron radiation.\footnote{This assumption is valid for magnetic
fields larger than $\sim 1 \mu$G, which is typically the case in
PWNe. We note, however, that at late times (ages of order 10 kyr,
as suggested for \pwn), when the magnetic field
decreases due to expansion, IC losses can begin to be significant.}  The
time evolution of the pulsar spin-down is determined by 
\begin{equation} \dot{E}(t) = \dot{E_0} \left(1 +
t/\tau_0\right)^{-\frac{b+1}{b-1}} \end{equation} 
where $\tau_0$ is a characteristic spin-down timescale for the
pulsar and $b$ is the pulsar braking index. Pulsations from the
putative pulsar have not yet been detected. To estimate the current
spindown power, we use the empirical relationship for $\dot{E}/L_x$
obtained by Possenti et al.  (2002), which yields $\dot{E} = 4
\times 10^{36} {\rm\ erg\ s}^{-1}$.  Based on the observed scatter
in the $\dot{E}/L_x$, the uncertainty on this estimated value may
be a factor of 10 or larger.  The maximum particle energy is
assumed to be limited by the condition that the particle gyroradii
do not exceed the termination shock radius (de Jager \& Harding
1992).  The lower limit on the particle energy is a free parameter,
and the overall normalization is set by the wind
magnetization parameter $\sigma$, equal to the ratio of power in
particles to that in the Poynting flux. The magnetic field in the
nebula is assumed to evolve as 
\begin{equation} B(t) = B_0/\left[
1 + (t/\tau_0)^\alpha\right] \end{equation} 
where $\alpha$ is a
free parameter.

\begin{figure}[t]
\epsscale{1.2}
\plotone{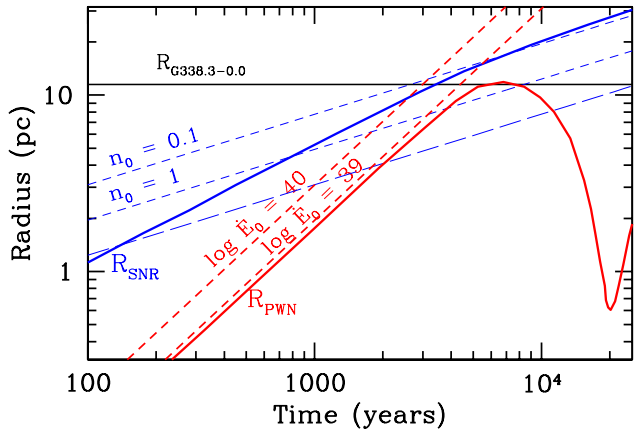}
\epsscale{1.0}
\caption{
Time evolution of the SNR and PWN radii for a range of values for
the ambient density and initial spin-down power of the pulsar. The
solid curves correspond to models from Gelfand et al. (2009) using
$\dot{E_0} = 10^{40} {\rm\ erg\ s}^{-1}$, $M_{ej} = 8 M_\odot$, $n_0
= 0.1 {\rm\ cm}^{-3}$, and $E_{51} = 1$.  See text description for
details.
}
\end{figure}

The broadband emission model results are shown in Figure 5 where
we plot the \fermi\ and H.E.S.S spectra along with the radio upper
limit from GMRT observations (Giacani et al. 2008) and spectral
confidence bands derived from Chandra (Lemiere et al. 2009).  The
black curves represent the model prediction for the synchrotron
(left) and inverse-Compton (right) emission that best describes the
X-ray and TeV $\gamma$-ray spectra, similar to results from Lemiere
et al. (2009); the parameters for the model, which were adjusted ``by
hand'' to provide good agreement with the radio, X-ray, and TeV
$\gamma$-ray data, as well as the inferred size of the PWN, 
are summarized in the
caption.  As seen in Figure 5, this model significantly underpredicts
the observed \fermi-LAT emission.

\begin{figure}[t]
\epsscale{1.2}
\plotone{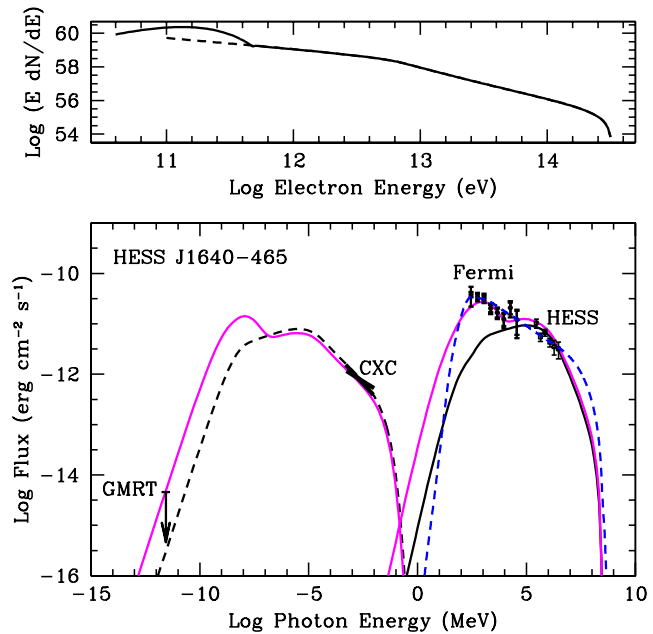}
\epsscale{1.0}
\caption{
Electron spectrum (upper) and broadband emission model (lower) for
\pwn\ assuming the evolutionary history described in the text. The
black curves represent a PWN with an age $T = 10$~kyr, and $B(T) =
5 \mu$G, assuming $\dot{E}(T) = 4 \times 10^{36} {\rm\ erg\ s}^{-1}$
and an injection spectrum with $\sigma = 10^{-3}$, $\gamma = 2.5$,
$\tau_0 = 500$~yr, and $E_{\rm min} = 115$~GeV. The magnetic field
evolution is characterized by $\alpha = 0.65$. The magenta curves
represent the scenario with a low-energy Maxwellian electron component
replacing the low-energy portion of the electron power-law spectrum.
The mean temperature for the IR and optical photon fields are 15~K
and 5000~K, respectively, and the energy densities relative to the
CMB are 4 and 1.15. The dashed curve in the upper panel represents
the truncated portion of the power law that was replaced by a
Maxwellian. The dashed blue curve in the lower panel represents a
model for which all of the $\gamma$-ray emission results from pion
decay.
}
\end{figure}

As discussed above, our spectral fits can formally accommodate up
to about $\sim 20\%$ of the observed flux in a pulsar-like component
characterized by a power law with an exponential cutoff energy
between 1 and 8 GeV. This corresponds to an energy flux (above
100~MeV) of $\sim 3.8 \times 10^{-11}{\rm\ erg\ cm}^{-2}{\rm\
s}^{-1}$. For the spin-down power suggested by its X-ray luminosity,
the available energy flux from the pulsar that powers \pwn\ is $\sim
3.3 \times 10^{-10} d_{10}^{-2}{\rm\ erg\ cm}^{-2}{\rm\ s}^{-1}$.
Thus, as much as 10\% of its spin-down could conceivably be
contributing directly to $\gamma$-rays in the LAT band. There are
known radio pulsars within the field of 1FGL~J1640.8$-$4634 as well.
PSR~J1637$-$4642, for example, is located within $\sim 37$~arcmin
and has a total energy flux of $2 \times 10^{-10} {\rm\ erg\
cm}^{-2}{\rm\ s}^{-1}$ given its estimated distance and measured
spin-down. Thus, while our spectral fits do not require a pulsar-like
component, it it quite feasible that one or more of these sources
contributes as much as 20\% of the observed flux. Inspection of
Figure 5 makes it clear, however, that the remaining LAT emission
still vastly exceeds the predicted flux from \pwn.

We note that simple power-law injection models for Vela X, another
evolved PWN, fail to reproduce the observed broadband spectrum
(LaMassa, Slane, \& de Jager 2009). The presence of an excess
population of low-energy electrons has been suggested, and models
for the inverse-Compton scattering of photons from this population
predict an excess of $\gamma$-rays in the GeV range (de Jager,
Slane, \& LaMassa 2009). This excess has been confirmed with
observations by both AGILE (Pellizzoni et al. 2010) and \fermi\
(Abdo et al. 2010b).  Motivated by this result, we modified the
evolved power law spectrum from our model for \pwn\ by truncating
the lower end of the power law and adding a distinct low-energy
component. Based on results from simulations of shock acceleration
(Spitkovsky 2008), we chose a Maxwellian distribution for this
population.  Our resulting (ad hoc) particle spectrum is shown in
the upper panel in Figure 5, and the resulting broadband emission
is shown in the magenta curves in the lower panel. Here we have
adjusted the normalization of the Maxwellian to reproduce the
emission in the \fermi-LAT band, which is produced primarily by
upscattered infrared (IR) photons from local dust. The energy density
and mean temperature of the IR photon field was adjusted slightly
to improve the agreement between the data and the model, but the
values (listed in the caption) are within reasonable expectations
(see, e.g., Strong et al. 2000). We find a mean value of $\gamma
\approx 2 \times 10^5$ for the electrons in the Maxwellian component,
and roughly 5\% of the total electron energy in the power law tail,
consistent with results from particle-in-cell simulations.\footnote{A.
Spitkovsky, private communication.} The associated pair multiplicity
relative to the integrated Goldreich-Julian injection rate is of
order $10^6$, similar to that inferred for the Crab Nebula as well
as several other PWNe (see Bucciantini et al. 2010).  Recent work
by Fang \& Zhang (2010) uses a similar input distribution to
successfully model the emission for several PWNe including \pwn.
However, their results for \pwn\ underpredict the observed GeV
emission from this source, apparently due to use of a slightly lower
bulk Lorentz factor and a larger fraction of the total energy in
the power law tail than we have used in this analysis.

We note that the Maxwellian shape for the low-energy electron
population is not unique. Indeed even a very narrow Gaussian
distribution can produce the GeV $\gamma$-ray emission without
exceeding the radio upper limit for the synchrotron emission. For
any of these models, the total energy in electrons requires a larger
initial spin-down period than assumed for the simple power-law
injection models -- by a factor of roughly 3 for the adopted
Maxwellian distribution. This is within the uncertainty in our
assumed value based on scaling from the X-ray luminosity of the
PWN. It is also important to note that while our proposed model is
consistent with the observed properties of this system, there are
potential degeneracies in the effects of different model parameters,
meaning that the proposed scenario is not necessarily unique.

An alternative scenario for the $\gamma$-ray emission is that it
arises from the SNR itself, and not the PWN.  Nonthermal bremsstrahlung
has been suggested as a mechanism for the production of $\gamma$-rays
in SNRs (e.g. Bykov et al. 2000). This process can be dominant for
electron-to-proton ratios of order 1.  However, the value typical
of local cosmic rays is closer to $10^{-2}$, and even smaller values
appear favored in models for $\gamma$-ray emission from SNRs (e.g.
Morlino et al. 2009, Zirakashvili \& Aharonian 2010, Ellison et al.
2010), so that this process is typically not dominant.  The dashed
blue curve in Figure 5 represents a model for the emission from the
collision of protons accelerated in the SNR with ambient material,
leading to $\gamma$-rays from the production and subsequent decay
of neutral pions. The $\gamma$-ray spectrum is calculated based on
Kamae et al. (2006) using a scaling factor of 1.85 for helium and
heavy nuclei (Mori 2009), and we have used a power law distribution
of protons with $dN_p/dE \propto E^{-2.4}$ to best reproduce the
observed spectrum. Assuming a shock compression ratio of 4 and that
25\% of the total supernova energy appears in the form of relativistic
protons, an ambient density $n_0 \approx 100 {\rm\ cm}^{-3}$ is
required to produce the model shown in Figure 5. This is much higher
than can be accommodated for the observed size of the SNR and the
lack of observed thermal X-ray emission from the SNR.  Such high
densities are found in dense molecular clouds, suggesting that the
$\gamma$-rays could be produced by particles that stream away to
interact with high-density material outside the SNR. However, only
the most energetic particles can escape the acceleration region,
which is in conflict with the proton spectrum we require to match
the data. Moreover, the observed TeV emission appears to originate
from within the SNR boundaries, making such an escaping-particle
scenario appear problematic.  Based on this, along with the lack
of a spectral cutoff that might suggest emission from a central
pulsar, we conclude that the GeV $\gamma$-ray emission most likely
arises from the PWN.

It is well-known that a distinct low-energy electron population
resides in the Crab Nebula, although the origin is not well-understood.
Atoyan (1999) has suggested that the spin-down timescale $\tau_0$
(see Equation 1) may itself be time-dependent, resulting in a large
energy input in the earliest epoch of pulsar spin-down, when
significant synchrotron and adiabatic losses would result in rapid
cooling of the electrons. Studies of the broadband emission from
3C~58 indicate an injection spectrum that differs from a pure power
law as well (Slane et al. 2008). As noted above, observations of
Vela~X appear to require an excess of low-energy electrons which
may be a relic population or could have been produced through rapid
synchrotron losses associated with the increased magnetic field
strength during the reverse-shock crushing stage.  More complete
modeling of the broadband emission of \pwn, accounting for the full
dynamical evolution of the system, including the effects of the
reverse shock on the PWN, is required to assess the overall energetics
and underlying particle spectrum more completely, but is beyond the
scope of the results we report here.

\section{Conclusions}

Broadband studies of \pwn\ have identified this source as a likely
PWN, with X-ray observations providing images of an extended nebula
as well as the putative pulsar powering the system. Modeling of the
PWN evolution based on inferred parameters of the pulsar imply
detectable emission in the GeV $\gamma$-ray band, and our investigations
of the \fermi-LAT observations of this region reveal clear evidence
of such emission, consistent with the source 1FGL~J1640.8$-$4634. 
The flux and spectrum we derive are consistent with that
of the previously-identified source 3EG~J1639$-$4702, and the much-improved
position makes it likely that the emission arises from \pwn. The
lack of a spectral cutoff rules out an association with emission
directly from the pulsar that powers the nebula, and the flux is
inconsistent with $\gamma$-rays from \snr, in which \pwn\ 
resides.

We have investigated the radio, X-ray, and $\gamma$-ray emission
from \pwn\ in the context of a simple one-zone model in which power
is injected into a nebula at a time-dependent rate consistent with
the current observed X-ray emission from the system. We find that
models constrained by the observed size of the associated SNR, as
well as limits on the size of the PWN, require an approximate age
of 10~kyr and a current magnetic field strength of only $\sim 4
\mu$G, consistent with expectations for the late-phase evolution
of a PWN. The conditions in such an evolved PWN yield a considerably
higher $\gamma$-ray flux, relative to the X-ray flux, than in younger
systems where the higher magnetic field results in significant
synchrotron radiation.

The observed \fermi-LAT emission from \pwn\ significantly exceeds
that predicted by our broadband models. We propose that the excess
emission is a signature of a distinct population of low-energy
electrons similar to that inferred from studies of the Crab Nebula
and Vela~X, although the nature of this electron component is not
well constrained. Deeper radio observations are needed to place
stronger constraints on this population. Sensitive searches for
pulsations from the central pulsar are of particular importance to
constrain the spin-down properties of the system, which can only
be very roughly constrained at present. There has been considerable
success in identifying such pulsars with the \fermi-LAT, but the
lack of an obvious pulsar-like spectrum in \pwn\ may argue for more
likely success with deep radio timing searches.

\acknowledgments

The work presented here was supported in part by NASA Contract
NAS8-39073 (POS) and \fermi\ Grant NNX09AT68G.  JDG is supported
by an NSF Astronomy and Astrophysics Postdoctoral Fellowship under
award AST-0702957.  POS, SF, and YU are grateful to the KITP in
Santa Barbara, where elements of the work presented here were first
discussed during a KITP program. The authors would like to thank
Don Ellison, Luke Drury, Felix Aharonian, and David Smith for helpful
discussions during the preparation of this manuscript.

The $Fermi$ LAT Collaboration acknowledges support from a number
of agencies and institutes for both development and the operation
of the LAT as well as scientific data analysis. These include NASA
and DOE in the United States, CEA/Irfu and IN2P3/CNRS in France,
ASI and INFN in Italy, MEXT, KEK, and JAXA in Japan, and the
K.~A.~Wallenberg Foundation, the Swedish Research Council and the
National Space Board in Sweden. Additional support from INAF in
Italy and CNES in France for science analysis during the operations
phase is also gratefully acknowledged.


\end{document}